\documentclass[aps,amsfonts,showpacs]{revtex4-1}
\usepackage{amsmath}
\usepackage{graphicx}

\usepackage{natbib}
\usepackage{psfrag}
\usepackage{amssymb}
\usepackage{color}
\usepackage{hyperref}
\usepackage{epstopdf}%

\DeclareGraphicsExtensions{.eps}

\begin{document}
\def\be{\begin{equation}}
\def\ee{\end{equation}}
\def\beq{\begin{equation}}
\def\eeq{\end{equation}}
\def\bea{\begin{eqnarray}}
\def\eea{\end{eqnarray}}
\def\bml{\begin{mathletters}}
\def\eml{\end{mathletters}}
\def\b{\bullet}
\def\eqn#1{(~\ref{eq:#1}~)}
\def\no{\nonumber}
\def\av#1{{\langle  #1 \rangle}}
\def\m{{\rm{min}}}
\def\M{{\rm{max}}}
\newcommand{\ds}{\displaystyle}
\newcommand{\tc}{\textcolor}
\newcommand{\TODO}[2][To do: ]{{\textcolor{red}{\textbf{#1#2}}}}

\title{Real eigenvalues of non-Gaussian random matrices and their products}    
\author{Sajna Hameed \footnote{e-mail:hamee007@ umn.edu }}
\affiliation{School of Physics and Astronomy, University of Minnesota, Minneapolis, MN 55455. U.S.A..}
\author{Kavita Jain \footnote{e-mail: jain@jncasr.ac.in}}
\affiliation{Theoretical Sciences Unit,
Jawaharlal Nehru Centre for Advanced Scientific Research,\\
Jakkur P.O., Bangalore 560064, India.}
\author{Arul Lakshminarayan \footnote{e-mail: arul@physics.iitm.ac.in}}
\affiliation{Department of Physics, Indian Institute of Technology Madras, Chennai, 600036, India}  
\affiliation{Max Planck Institute for the Physics of Complex Systems, 01187 Dresden, Germany}

\widetext
\date{\today}

\begin{abstract}
We study the properties of the eigenvalues of real random
matrices and their products. It is known that when the matrix elements
are Gaussian-distributed independent random variables, the fraction of
real eigenvalues tends to
unity as the number of matrices in the product increases. Here we
present numerical evidence that this phenomenon is robust with respect
to the probability distribution of matrix elements, and is therefore a
general property that merits detailed investigation. 
Since the elements of the product matrix are no longer distributed as
those of the single matrix nor they remain independent random
variables, we study the role of these two factors in detail. We study
numerically the properties of the Hadamard (or Schur) product of
matrices and also the product of matrices whose entries are {\it
  independent} but have the same
marginal distribution as that of normal products of matrices, and find
that under repeated multiplication, the probability of all eigenvalues
to be real increases in both cases, but saturates to a constant below
unity showing
that the correlations amongst the matrix elements are responsible for
the approach to one. To investigate the role of the non-normal
nature of the probability distributions, we present a thorough
analytical treatment of the $2 \times 2$ single matrix for
several standard distributions. Within the
class of smooth distributions with zero mean and finite variance, our
results indicate that the Gaussian distribution has the 
maximum probability of real eigenvalues, but the Cauchy distribution
characterised by infinite variance is found to have a 
larger probability of real eigenvalues than the normal. We also find
that for the two-dimensional single matrices, the probability of real
eigenvalues lies in the range $[5/8,7/8]$.

\end{abstract}
\maketitle
%


\section{Introduction}

The problem of the number of real roots of algebraic equations has a
long history, but continues to attract attention, from the early
works of Littlewood and Offord \cite{Littlewood1939}
to more recent developments \cite{Edelman1995,Nguyen2014} (see
the latter article for more related history and references). 
Mark Kac, in a seminal paper, proved that the expected number of real
zeros of polynomials (of order $N$) whose coefficients are chosen from
a normal distribution of zero mean is $\sim \frac{2}{\pi}\ln{N}$
\cite{Kac1943}. Others substantially extended these results and showed 
universality in that this is the leading order behavior, irrespective of the
underlying distribution as long as they are zero-centered and have a
finite variance \cite{Stevens1969}. Logan and Shepp
\cite{LoganShepp1966} studied the same when the coefficients are
Cauchy distributed (and hence have infinite variance, and undefined
average) the leading behavior goes as $c \ln N$, 
with $c \approx 0.7413$ which is larger than $2/\pi\approx 0.6366$, and
hence there are more real zeros in the Cauchy case than when the
variance is finite. Thus, the number of real roots of a random
polynomial are generally quite small.

More recently, several works have explored the fraction of real
eigenvalues for $n \times n$ matrices drawn from the real Ginibre
ensemble. This ensemble consists of matrices whose elements are
independently drawn from a normal distribution such as $N(0,1)$. For
such matrices, it has been shown analytically that the expected number
of real eigenvalues $E _n$ and the probability that {\it all}
eigenvalues are real $P_{n,n}$ are given by
\cite{Edelman1994,Edelman1997,Kanzieper2005} 
\begin{equation}
\lim_{n \to \infty} \frac{E_n}{\sqrt{n}}= \sqrt{\frac{2}{\pi}},\;\; P_{n,n} = 2^{-n(n-1)/4}.
\label{eq:GinExpReal}
\end{equation}
Thus the probability that all eigenvalues are real tends to zero as the
matrix dimension increases, although the expected number of them
increases algebraically.

Products of random matrices have also been studied, at least,
since the classic work of Furstenberg and Kesten
\cite{Furstenberg1960} (see, \cite{Bougerolbook,Crisantibook} for a
discussion of subsequent work and some applications). The study of the
spectra and singular values of the products of random matrices is
currently a very active area of research
\cite{Roga2011,Burda2012}, and recent progress has been reviewed in
\cite{Akemann2015}. 
When studying a problem related to the measure of ``optimally
entangled" states of two qubits \cite{Arul2011}, the problem of the
number of real eigenvalues of a product of matrices came up. It was
shown in \cite{Arul2013} that the probability of all eigenvalues being
real for the product of two, $2\times 2$ matrices from the real
Ginibre ensemble is $\pi/4$. Somewhat surprisingly then there is a
lesser 
probability that all eigenvalues are real for a single 
matrix ($1/\sqrt{2}$, from Eq.~(\ref{eq:GinExpReal})) than for a
product of two. This observation motivated  numerical explorations in
\cite{Arul2013} which showed that this probability tends to $1$ as the
number of matrices in the product increases. Numerical results showing
that this is true for higher dimensions  was also discussed
therein. This result has been proven analytically by Forrester
\cite{Forrester2014}, who calculated $P_{n,n}$ explicitly in terms of
Meijer G-functions for square matrices, and generalised by Ipsen
\cite{Ipsen2015} to rectangular matrices.

Most of the previous works on the properties of the eigenvalues have studied
Gaussian matrices with independent, identically distributed (i.i.d.)
entries. Exceptionally,
Edelman, Kostlan and Shub \cite{Edelman1994} presented some numerical
results indicating universality  for the expected value $E_n$ of real
eigenvalues of $n \times n$ real matrices. However, to the best of our
knowledge, there is no study exploring such extensions to the products of
matrices. Thus,  two conditions are relaxed: first, non-Gaussian i.i.d. elements are taken as entries of the matrices, and second, we consider products of  such independent matrices. In this case, the entries of the product matrix are naturally correlated in a complex manner and their distribution is, in general, anyway non-Gaussian. 

Motivations for studying products of random matrices
are well known in physics, and range from the study of localization in random media where transfer matrices are multiplied to dynamical systems where products of 
local stability matrices determine the Lyapunov exponents \cite{Crisantibook}. Applications for counting number of real eigenvalues of random matrices, especially of products, is of more recent provenance. The real eigenvalues of a class of random matrices indicate topologically protected level crossings at the Fermi energy in the bound states of a Josephson junction \cite{Beenakker2013}. While the relevant object is a single matrix and not a product, it is conceivable that this provides a context for further investigation. In the context of entanglement of two qubits \cite{Arul2013}, the real eigenvalues of products of two matrices 
were relevant in finding the measure of optimally entangled states \cite{Arul2011}. Different distributions of matrix elements would result in different sampling of states on the Hilbert space. 
Finally, it is worth pointing out that the fact that real eigenvalues dominate the spectrum of products of matrices implies that generic orbits of dynamical systems are not of the complex unstable variety.
Complex instability can occur in Hamiltonian  systems with more than two degrees of freedom \cite{Contopoulos1994}, when exponential instability is combined with rotational action in phase space. However, as a consequence of discussions in this paper, it seems unlikely that they would generically arise for long orbits, as they depend on eigenvalues of products of Jacobian matrices being complex.

To begin with, we ask: how do non-normal distributions affect the probability of real eigenvalues of a single matrix? In fact, a major part of
this paper is concerned simply with $2 \times 2 $ matrices and
various families of the distributions of matrix elements, and the
probability that 
the eigenvalues are real is calculated exactly in many cases in Sec.~\ref{single}. In particular, we find the lower and upper bounds on
the probability that both eigenvalues are real. In the
family of symmetrized Gamma 
distributions, the range of probabilities is shown to be
$[5/8,7/8]$. The upper bound is associated with the case when the
matrix elements have a large weight at the origin and the lower bound in the
opposite case when the maximum is away from the origin. In fact, these
same bounds are obtained in a class of truncated distributions thus 
indicating that the probability of real eigenvalues depends crucially
on whether  the distributions are concentrated near the origin or away
from it. {The lower bound of $5/8$ is shown to be a tight one by proving this in general. The upper bound of $7/8$ still remains specific to these distributions, however we do believe that this is widely applicable as well.}

We also find that the other features of the distributions of the
matrix elements that play a significant role in determining the probability of real eigenvalues are the degree of smoothness and the
existence of finite moments. It is noteworthy that these (apart
from zero mean) are also the crucial features for universality in the
case of random polynomials \cite{Edelman1995}. 
Given that the distributions are smooth and have finite moments, we
tentatively propose that the normal distribution is the one with the
maximum probability of finding real eigenvalues. If true, this
provides a further unique characterization of the normal. 

Besides the
bounds, we also observe a hierarchy for the fraction of real
eigenvalues for matrices constructed from these
distributions. Specifically, for some commonly occurring
distributions, we find that the eigenvalues are more
real when the matrix elements are chosen from Cauchy distribution than
Laplace, Gaussian or uniform distributions, which are arranged in the
increasing order of the 
decay. However, we must mention that such hierarchy patterns are not
simply determined by the tail behavior of the distribution, and the
trends appear to be more complex as explained in the following section. 

In Sec.~\ref{secproduct}, we numerically study the properties of the
eigenvalues of the matrix 
obtained after taking product of several matrices, and find that the
increase in the probability to one holds for several other 
distributions. In particular, we explore the zero-centered uniform
distribution, the symmetric exponential (Laplace) distribution and
Cauchy distribution, as simple representative ones.  
The hierarchy mentioned above is seen to hold even after the multiplication of
independent random matrices. 
The expected number of real eigenvalues
is also studied which tends to the matrix dimension
exponentially fast (with the number of matrices in the product), 
although the rate of approach is distribution-dependent.  

It is unclear as to why the eigenvalues tend to become real when
multiplying random structureless matrices. In an attempt at seeing how
much this has to do with the act of multiplication, we study
numerically the probability of real eigenvalues in the case of
Hadamard (or Schur) products where the elements  are simply the products
of the corresponding elements of the 
multiplying matrices. Of course, in this case, the matrix elements
remain uncorrelated. We find numerically that the probability of real eigenvalues 
increases with the number of matrices in the product, but tends to
saturate at a value smaller than $1$. Numerical evidence that this
approach is a power law  is also provided. This then highlights that the
{\it correlations} built up in the process of (usual) matrix multiplication
are responsible for the phenomenon that the fraction of
real eigenvalues tends to one.

\section{Real eigenvalues of a single $2\times 2$ matrix}
\label{single}

Let $P_{n,k}^{(K)}$ denote the probability that a product of $K$ $n
\times n$ random matrices has $k$ real eigenvalues. 
In this section, we study the simplest case, viz., the probability
$P_{2,2}^{(1)}$ that all 
the eigenvalues of a $2 \times 2$ matrix are real. We assume that the matrix
elements are i.i.d. random variables chosen from the distribution $p(x)$ with
support on the interval $[-u,u]$, where $u$ is finite for bounded
distributions and infinity for unbounded ones, and that the
distribution $p(x)$ is symmetric about the origin 
as a result of which the  mean of the probability distribution is
guaranteed to be zero. We first give analytical results for the
probability $P_{2,2}^{(1)}$, which we study for many probability
distributions, and then present some numerical results for the more general
quantity $P_{n,k}^{(K)}$ in the following section.


Consider a $2\times 2$ matrix with i.i.d. elements defined as
\begin{center}
$\left[\begin{array}{cccc}
v & x \\
y & w 
\end{array}\right].$
\end{center}
As the discriminant $\mbox{tr}^2 -4 
\det$ must be nonnegative for real eigenvalues, the probability that
all eigenvalues are real is  
\begin{equation}
\begin{split}
P_{2,2}^{(1)} =
\int^{u}_{-u}\int^{u}_{-u}\int^{u}_{-u}\int^{u}_{-u} \Theta\left[(v-w)^2
  + 4xy\right] ~p(v) p(w) p(x) p(y)~ dv dw dx dy.
\end{split}
\label{eq:3}
\end{equation}
Let the probability distribution of $z = v - w$ be $q(z)$ with $-2 u
\le z \le 2u$.  
Observe then that the probability of real eigenvalues can be written as 
\be
P_{2,2}^{(1)} =
\frac{1}{2} +
4\int^{0}_{-u} dx \int^{u}_{0} dy \int^{2u}_{0} dz ~q(z)
p(x) p(y) \Theta\left[z^2 +
4xy\right]. 
\label{eq:Prob2Intgl}
\ee
Here the even symmetry of the distributions is used, and also that when
the signs of both $x$ and $y$ are the same, the constraint imposed by
the Heaviside function is trivially satisfied. In the above equation,
the distribution $q(z)$ is given by the convolution 
\be
q(z)=\int_{-u}^{u} ~p(v) ~p(w) ~\delta(z-(v-w)) ~dv dw  =\int_{-u}^{u} ~p(v) p(v-z) ~dv. 
\label{convol}
\ee
We note that this is indeed the usual convolution with $z \rightarrow
-z$ due to the symmetry of $p(x)$. If $z \in [-2 u ,2 u]$ the above
expression for $q(z)$ is valid, else it is zero.

Equation (\ref{eq:Prob2Intgl}) shows that the probability of
both eigenvalues being real is at least one half for any probability
distribution. The integral in
Eq.~(\ref{eq:Prob2Intgl}) can be simplified if we use the convolution
as  
\be
P_{2,2}^{(1)} =
\frac{1}{2} +
4\int^{u}_{0} dx \int^{u}_{0} dy \int^{2u}_{2 \sqrt{x y}} dz ~q(z)
p(x) p(y)= 1-4\int^{u}_{0} dx \int^{u}_{0} dy \int_{0}^{2 \sqrt{x y}} dz ~q(z)
p(x) p(y).
\label{eq:Prob2IntglConv}
\ee
The last equality follows from the fact that the convolution is itself
a symmetric normalized distribution as $\int_{-\infty}^{\infty} q(z)
dz=1$. Thus the evaluation of the probability of real eigenvalues
reduces to evaluating the triple integral above. 

It is also useful to write the distribution $q(z)$ as 
\be
q(z)= \frac{1}{2 \pi} \int_{-\infty}^\infty dk ~e^{i k z} ~|{\tilde p}(k)|^2,\;\;  {\tilde p}(k)= \int_{-u}^u p(x) ~e^{-i k x} ~dx.
\label{FT22}
\ee
Here ${\tilde p}(k)$ is the characteristic function (or Fourier transform) of the probability distribution $p(x)$.
Next consider the integral defined as 
 \begin{equation}
 I(\alpha) = \int^{0}_{-u} dx \int^{u}_{0} dy \int^{2u}_{0} dz ~ q(z)
 p(x) p(y) \Theta\left[z^2 + \alpha x y\right].
 \label{eq:5}
 \end{equation}
 The value of interest is $I(4)$. Towards this end, differentiating $I(\alpha)$ with respect to (w.r.t.) $\alpha$, and performing the 
 integral over the resulting delta function leads to  
 \be
 \frac{\partial I}{\partial \alpha} 
 =\int^{\infty}_{-\infty} d\omega \int^{u}_{0} dx \int^{u}_{0} dy
 \frac{xy}{2\sqrt{\alpha xy}}\frac{\left|{\tilde
     p}(\omega)\right|^2}{2\pi}e^{i\omega\sqrt{\alpha xy}}p(x)p(y), 
 \label{eq:8}
 \ee
 where ${\tilde p}(\omega)$ is given by Eq.~(\ref{FT22}). Noting 
 that the above equation is valid only for $\alpha \in (0,4)$, we 
 integrate the last expression w.r.t. $\alpha$ from $0$ to $4$, and
 use $I(0)=1/8$ to get $I(4)$ and finally the probability that both the eigenvalues are real as 
 \begin{equation}
 P_{2,2}^{(1)} = 1 -
 \frac{4}{\pi}\int^{\infty}_{0} d\omega \int^{u}_{0} dx \int^{u}_{0} dy \left|{\tilde
   p}(\omega)\right|^2\frac{\sin(2\omega\sqrt{xy})}{\omega} p(x)p(y).
 \label{eq:Prob2Charact}
 \end{equation}
 Below we will apply the result in either Eq.~(\ref{eq:Prob2IntglConv}) or  Eq.~(\ref{eq:Prob2Charact}) to various choices of
 distribution $p(x)$.  
Before turning to explicit calculations, we note that the unbounded
distributions have the following scale invariance property. Consider a
division of the random matrix elements by a nonzero constant $b$.
If $x'=x/b$, then $\int_{-u'}^{u'} dx'p'(x')=1$ where $p'(x')=b p(x)$
and $u'=u/b$. From Eq.~(\ref{eq:3}), it is seen 
  that the probability $P_{2,2}^{(1)}$ is not affected by this scaling for
  unbounded functions, but for bounded ones, the 
  limits should be redefined. While the above is
an elementary analysis, we are not aware of it being discussed
before. 

Under conditions of smoothness (in the sense that all the
derivatives exist at least in intervals) and symmetry of $p(x)$, it is 
interesting to enquire about the maximum value or upper bounds of
$P_{2,2}^{(1)}$, for instance. For the lower bound, we have already
stated that the probability of all real eigenvalues is at least one
half. A tighter lower bound is however $5/8$, a proof of which was suggested to
us by an anonymous referee which we discuss now. The probability of eigenvalues being real is 
\beq
\mathbb{P}[(v-w)^2+4 xy >0]=\frac{1}{2}\mathbb{P}[(v-w)^2+4 xy >0| xy>0]+\frac{1}{2}\mathbb{P}[(v-w)^2+4 xy >0| xy<0],
\eeq
where the factors of $1/2$ arise as probability that $xy>0$ or
otherwise. However the first conditional probability is $1$, and therefore
\bea
\mathbb{P}\left[(v-w)^2+4 xy
    >0\right] &=& \frac{1}{2}+\frac{1}{2}\mathbb{P}[(v-w)^2-4 xy >0| xy>0] \\
&=& \frac{1}{2}+\frac{1}{4}\mathbb{P}[(v+w)^2-4 v w -4 xy >0| xy>0, v
  w >0] \nonumber \\
  &+&\frac{1}{4}\mathbb{P}[(v+w)^2-4 v w -4 xy >0| xy>0, v w
  <0] \\
&=& \frac{1}{2}+\frac{1}{4}\mathbb{P}[(v+w)^2-4 v w -4 xy >0| xy>0,
  v w >0] \nonumber \\
&+& \frac{1}{8}\mathbb{P}[(v+w)^2-4 v w -4 xy >0| xy>0, v w
  <0,|v w|< xy]+\frac{1}{8} \\
&\geq& \frac{5}{8}.
\eea
Here, in the second equality, further conditions on the sign of $v w$
are used with equal probabilities for either occurrence (from the
symmetry and independence of the distributions). In the third
equality, a further certainty is carved out by using the possibility that when
$|vw|>xy>0$ (occurence  probability  being $1/2$),  the discriminant
is certainly positive. However our attempts at a similar approach for the
upper bound were not successful. 

Although one can
obtain get some insight into the bounds, for what kind of distributions 
these bounds actually occur is discussed below. 


\subsection{Bounded distributions}

\subsubsection{Symmetric Beta distribution}

Consider the symmetric Beta distribution with zero mean defined as 
\be
p_{\mu,\nu}(x)=\frac{\Gamma(\mu+\nu+2)}{2 \Gamma(1+\mu) \Gamma(1+\nu)}
~ (1-|x|)^\mu ~ |x|^\nu ~,~\mu, \nu > -1
\label{bddef}
\ee
with support on the interval $[-1,1]$. This is a two-parameter 
symmetric family which can be nonsmooth at $x=0$. 
For $\mu=\nu=0$, the above distribution reduces to a uniform distribution,
while $\mu=1,\nu=0$ and $\mu=0, \nu=1$ correspond to tent-shaped or
V-shaped distribution respectively.


\begin{center}
{\bf Case: ${\mathbf \mu=0}$}
\end{center}

\begin{table}[t]
\begin{center}
\begin{tabular}{|c|c|}
\hline
 Distribution $p_{0,\nu}(x)\sim |x|^{\nu}$ & Probability $P_{2,2}^{(1)}$ \\
\hline 
$\nu=-4095/4096$ & 0.874959 \\
$\nu=-7/8$ & 0.849868 \\
$\nu=-1/2$ &  0.759836 \\
$\nu=0$ &  0.680556 \\ 
$\nu=1$ &  0.63709 \\
 $\nu=3/2$ & 0.632888 \\
 $\nu=2$ &  0.631023\\
$\nu=3$ & 0.62928 \\
$\nu=4$&  0.628361\\
$\nu=200$ & 0.625078\\
$\nu=400$ & 0.625039\\
\hline
\end{tabular}
\end{center}
\caption{Probability that both eigenvalues are real for a $2 \times 2$
  matrix with matrix elements chosen independently from the
  distribution in Eq.~(\ref{bddef}), with $\mu=0$. As $\nu$ approaches
  $-1$, the probability seems to limit to  $7/8$, while for $\nu \to
  \infty$, it tends to $5/8$ (see text for details).} 
\label{ta:summary2}
\end{table}

For the uniform distribution, the
convolution $q(z)=(2-|z|)/4$ on $[-2,2]$ and zero elsewhere. Using
Eq.~(\ref{eq:Prob2IntglConv}), we find that 
\be
P_{2,2}^{(1)}(\nu=0)= \frac{1}{2} +\int_{0}^{1} dx \int^{1}_{0} dy
\int^{2}_{\sqrt{4 x y}} dz ~q(z)  
= \frac{49}{72} = 0.680556. 
\label{eq:UniformProb} 
\ee
Thus the uniform distribution results in a {\it smaller} fraction of
real eigenvalues when compared to the normal distribution
\cite{Edelman1994}, but not by very much.

When $\mu=0$ and $\nu$ is nonzero and positive, the distribution
$p_{0,\nu}(x) \sim |x|^\nu  \Theta(1-|x|)$ is zero at the origin. As $\nu$ tends to
infinity, the weight of the distribution gets increasingly concentrated
at $\pm 1$ and as Table~\ref{ta:summary2} shows, the probability of real
eigenvalues decreases. The distribution is smooth when $\nu=2k$ is an even
integer. In this case, an analytical expression is possible for
arbitrary $k$, and given by Eq.~(\ref{eq:Probmodxpownu}) of Appendix~\ref{app_beta}. 
Numerical analysis of Eq.~(\ref{eq:Probmodxpownu}) for large $\nu$ strongly suggests that 
\be
P^{(1)}_{2,2}(0, \nu)=\frac{5}{8} + {\cal O}(\nu^{-1}), \;\; \nu \gg 1.
\label{eq:nupos}
\ee
An insight into the above result can be gained by the following 
heuristic arguments: If we consider the limiting distribution as the Bernoulli ensemble with only two choices 
of entries $\pm 1$ with equal probabilities, we get only $16$ matrices
in the ensemble of which $12$ have real eigenvalues.  
Thus it would seem that the ratio should have converged to $12/16=3/4$. 
However, there are $4$ cases where both the eigenvalues are exactly zero.
If we believe that in the case of continuous distributions these are
modified into cases with real and complex eigenvalues and equally, we
get $6$ cases of complex values and the fraction is consistent with
$5/8$ for real eigenvalues. 
 Another related heuristic argument, closer to the continuous distributions under consideration, starts with the distribution
 \beq
 p(x)=\frac{1}{2}(\delta(x+1)+\delta(x-1)),
 \eeq
 and applies Eq.~(\ref{eq:Prob2Charact}) with $\tilde{p}(\omega)=\cos(\omega)$, leading to  
 \[ P^{(1)}_{2,2}=
1-\frac{2}{\pi}\int_0^{\infty} \dfrac{\sin \omega \cos^3 \omega}{\omega} \, d \omega= \frac{5}{8}.
 \]
A proof of the assertion in
Eq.~(\ref{eq:nupos}) by analysing the double sums in 
Eq.~(\ref{eq:Probmodxpownu}) seems rather difficult to obtain, but it is
possible to tackle the Gamma distribution given by 
Eq.~(\ref{eq:gammadef}), for
which also the probability $p(x)$ is zero at the 
origin and increases algebraically away from it, and show that indeed
the limit probability is $5/8$, see Sec.~\ref{sec_gamma}.  Thus the lower bound derived above is a tight one.

We now turn to the case when $-1<\nu<0$ where the probability
distribution $p_{0,\nu}$ piles up
at the origin. As shown in Table \ref{ta:summary2}, the
probability of real eigenvalues increases as $\nu$ decreases towards
$-1$. It is obvious that if the matrix elements are chosen from a 
Dirac-delta distribution centred about zero, both the eigenvalues are 
definitely real. But whether the probability limits to something less
than one as $\nu$ approaches $-1$ is of natural 
interest. As shown in Appendix~\ref{app_beta}, we find that 
\be
P_{2,2}^{(1)} (0,\nu)= \frac{7}{8} - {\cal O}(1+\nu), ~~ \nu \to -1.
\ee

Thus it is interesting that the distribution $\sim |x|^\nu$ restricted
to an interval spans a range of behaviors for the probability of real
eigenvalues. As the matrix elements probability gets increasingly
piled up at the ends ($\pm 1$), the probability of real eigenvalues decreases and tends to $5/8$, while in the opposite case when the
elements are piled up around origin, the probability approaches the
maximum value of $7/8$.
  

\begin{center}
{\bf Case: $\mathbf \nu=0$}
\end{center}

\begin{table}[t]
\begin{center}
\begin{tabular}{|c|c|}
\hline
 Distribution $p_{\mu,0}(x) \sim (1-|x|)^{\mu}$ & Probability $P_{2,2}^{(1)}$ \\
\hline 
 $\mu=-1/2$ & 0.654534 \\ 
$\mu=1/2$ & 0.695759 \\
$\mu=3/4$ &0.70085 \\
$\mu=1$ &0.704854 \\
\hline
\end{tabular}
\end{center}
\caption{Probability that both eigenvalues are real for a $2 \times 2$
  matrix with matrix elements chosen independently from the distribution in Eq.~(\ref{bddef}) with $\nu=0$.  The probability lies
  in the interval [$5/8,11/15$] for $\mu \in (-1,\infty)$.}
\label{ta:summary1}
\end{table}

For $\nu=0$, the probability distribution $p_{\mu,\nu}(0) \neq 0$ and
we find that with
increasing $\mu$, the probability that the eigenvalues are real
increases, refer Table~\ref{ta:summary1}. 
Partial analytical results are possible. For example, for the case when $\mu=1, \nu=0$ (``tent" distribution), the convolution is 
\be
q(z)= \left\{ \begin{array}{ll}\frac{1}{6}(2-|z|)^3& 1 \le |z| \le 2\\ \frac{1}{6}(4-6 |z|^2 +3 |z|^3) & 0 \le |z| \le 1. \end{array} \right..
\label{eq:Convmu1}
\ee 
The resulting integral in Eq.~(\ref{eq:Prob2IntglConv}) can now be done using hyperbolic coordinates. This results in  
\be
P_{2,2}^{(1)} (\mu=1,\nu=0)=\frac{16143}{22400}-\frac{23 \ln 2}{1008} \approx 0.704854.
\label{eq:Probmu1}
\ee
 It is reasonable to expect that as $\mu$ increases (and $\nu=0$), the
probability increases to that when the elements are distributed
according to the Laplace distribution $\sim \exp(-|x|)$ which is
obtained as $\mu \rightarrow 
\infty$. We will deal with the Laplace distribution below, but state
here that the probability of real eigenvalues in this case is $11/15
\approx 0.733$.  However, for negative $\mu$ where the matrix elements
have a tendency to be close to $\pm 1$, from the discussion in the
last subsection, we expect the probability to approach $5/8$ as $\mu
\to -1$. Thus for the distribution $p_{\mu,0}(x)$, the probability of
real eigenvalues lies in the range $\left[5/8, 11/15 \right]$.


\subsubsection{A smooth family}
 \label{smoothbd}

Consider another class of smooth zero mean distributions bounded on the
interval $[-1,1]$ defined as 
\be
p_{\eta} (x)= \dfrac{\Gamma(\eta +\frac{3}{2})}{\sqrt{\pi} \Gamma(
  \eta +1)} (1-x^2)^{ \eta} ~,~\eta > -1.
\ee
Of course, for $\eta>0$, the distribution 
is continuous but not smooth at $\pm 1$, but this does not seem crucial. 
When $\eta=0$, $p_{\eta} (x)$ is uniform on the said interval, and as
$\eta \rightarrow \infty$ the distribution approaches the normal
distribution with the variance scaling as $1/\eta$. As discussed
above, the probability of real eigenvalues is independent of the
variance and therefore, we expect that the large $\eta$ 
value for this probability will coincide with the known result for the
normal distribution, namely, $1/\sqrt{2} \approx 0.707 \cdots$
\cite{Edelman1994} (also, see Sec.~\ref{sec_gauss}).  The probability
of real eigenvalues is calculated for some values of integer $\eta$
in Appendix~\ref{app_smooth}, and we find that the probability indeed
approaches $1/\sqrt{2}$ with increasing $\eta$. 

When $\eta$ is negative, the distribution
  $p(x)$ diverges at $x=\pm 1$, and from the discussion in the
  preceding subsection, we expect the probability of having real
  eigenvalues to approach $5/8$ as $\eta \to -1$. The case of $\eta=-1/2$ is that of the 
  arcsine distribution. The convolution with itself can be found and is a complete Elliptic
  integral. However, even if further analytic results seem to be hard,
  this enables a more accurate numerical estimate of the probability
  of real eigenvalues which is $\approx 0.662$. Further decreasing
  $\eta$ makes the numerical evaluation unstable, but results indicate
  a monotonic decrease in the probability.

  
\subsection{Distributions with infinite support}

\subsubsection{Gaussian}
\label{sec_gauss}

The case of normal or Gaussian distribution is the most studied and
there are general results
\cite{Edelman1994,Arul2013,Forrester2014}. We consider a zero centered, unit variance Gaussian distribution for the matrix elements given by 
\be 
p(x) = \frac{e^{-x^2/2}}{\sqrt{2\pi}};\;  x\in(-\infty,\infty).
\label{eq:Normal}
\ee 
The probability of both eigenvalues being real has already been shown,
using Eq.~(\ref{eq:Prob2Charact}), to be exactly equal to $1/\sqrt{2}$ in \cite{Arul2013}, and alternative derivations
naturally exist, see the works of Edelman \citep{Edelman1994} and
Forrester \cite{Forrester2014}. However to place it in the
context of this paper, we rederive the probability of real eigenvalues
in this case in Appendix~\ref{app_gauss}.  

It is well known that the normal or Gaussian distribution is singled
out in numerous ways, for example, as one that maximizes entropy for
a given mean and variance, or as the limit of sums of random
variables. In the context of the present work, it seems plausible that
the normal distribution is once again to be singled out as the
distribution of matrix elements  that maximizes the probability of
finding real eigenvalues among the class of {\it smooth, symmetric,
  and finite moments} distributions. To test this further, we have
looked at distributions $p(x)$ such as $\sim \exp(-x^4)$, $\sim
\exp(-x^2- r x^4)$ ($r>0)$ etc. and verified in all these cases that
the probability of real eigenvalues is indeed less than
$1/\sqrt{2}$. We advance this proposition tentatively based on
evidence gathered so far, and from results to be presented below.

\subsubsection{Gamma distribution}
\label{sec_gamma}

We next consider the {\it symmetrized} Gamma distribution defined as 
\be
p_\gamma(x)=\frac{1}{2 \Gamma(\gamma)} |x|^{\gamma-1} e^{-|x|} ~,~\gamma > 0.
\label{eq:gammadef}
\ee
where the scale for the exponential decay has been set to one. The
special case of Laplace distribution for which $\gamma=1$ and the
general case using the convolution route are discussed in
Appendix ~\ref{app_gamma}. The convolution
itself is given by 
\begin{equation}
q(z)=\dfrac{1}{4 \Gamma(2\gamma)} e^{-|z|} |z|^{2\gamma-1} + \dfrac{1}{2^{\gamma+\frac{1}{2}} \sqrt{\pi} \Gamma(\gamma)}|z|^{\gamma-\frac{1}{2}}K_{\gamma-\frac{1}{2}}(|z|),
\label{eq:ConvGamma2}
\end{equation}
where $K_{\nu}(z)$ is the modified Bessel function of the second kind
\cite{NIST:DLMF}. The results obtained by analysing the resulting
integrals are shown in Table~\ref{ta:gamma}, and we find that with
increasing $\gamma$, as the weight of the distribution at the origin
decreases, the probability $P_{2,2}^{(1)}$ also decreases.

The behavior of the probability of real eigenvalues as $\gamma$
approaches zero can be understood as follows. As $\gamma$ approaches
zero from  the positive side, the distribution and both the
terms in the convolution $q(z)$ 
diverge at the origin as $1/|z|$. The latter can be seen as 
   $K_{-1/2}(z)=K_{1/2}(z)=\sqrt{\pi/2}e^{-z}/\sqrt{z}$, and $\lim _{z
     \rightarrow 0} \Gamma(2 z)/\Gamma(z)=1/2$, as follows from the
   duplication formula for the Gamma function \cite{Abramowitz1964}. 
Since we expect that the divergence at the origin is all that matters,
the probability for real eigenvalues will converge to the case of the
bounded distribution $|x|^{\nu}$ already studied above, and therefore
the probability for real eigenvalues increases to $7/8$ as $\gamma \to
0$.

For increasing $\gamma >1$, the distribution itself is peaked away
from the origin and has two symmetric maxima at $\sim
\gamma$. 
The probability $P_{2,2}^{(1)}(\gamma)$ now decreases from the
$\gamma=1$ value and monotonically seems to approach $5/8$, the lower
limit for the bounded $|x|^{\nu}$ distributions as $\nu$
increased. Indeed the piling up of the probability of the matrix
elements at two symmetric ``walls" makes this plausible. To analyze
this further, we restrict attention to integer values of $\gamma$ and give an exact evaluation in terms of finite sums as 
\begin{equation}
P_{2,2}^{(1)}(\gamma)=\frac{1}{2}+A_1(\gamma)+A_2(\gamma), 
\label{eq:Probgamma}
\end{equation}
where $A_1$ and $A_2$ are given by Eq.~(\ref{A1A2}) 
and originate from the first and second terms of the convolution $q(z)$ in Eq.~(\ref{eq:ConvGamma2}) respectively. 
To find the limiting value
for the probability of real eigenvalues for large $\gamma$, we note
that like the distribution $p_\gamma(x)$, the first term in
the convolution $q(z)$ is
also peaked around $z \sim \pm 2 \gamma$, but the second term  peaks at
$z=0$ so that the two terms in the convolution have practically
  disjoint supports. As a consequence, for large $\gamma$, the
  dominant  contribution to the
probability in Eq.~(\ref{eq:Probgamma}) comes from $A_1(\gamma)$ which
represents the overlap between the distribution and the 
convolution, and $A_2(\gamma)$ can be neglected. 
As discussed in Appendix~\ref{app_gamma}, on analysing the sum
$A_1(\gamma)$, we get 
\be
P^{(1)}_{2,2}(\gamma)=\frac{5}{8} + \frac{1}{16 \sqrt{2 \pi \gamma}}+ {\cal O} (\gamma^{-1}), \;\; \gamma \gg 1.
\ee
In summary, for the symmetrized Gamma distributions, the probability of real eigenvalues also seems to be in the range $[5/8, 7/8]$. The distribution having a nonanalyticity at the origin leads to larger probability for real eigenvalues compared to the normal distribution and it seems comparable to the bounded power law distributions $|x|^{\nu}$ studied above except in the rates of convergence.

\begin{table}[t]
\begin{center}
\begin{tabular}{|c|c|}
\hline
 Distribution symmetric Gamma & Probability $P_{2,2}^{(1)}$ \\
\hline 
 $\gamma=1/4$ & 0.824051 \\
  $\gamma=1/2$ &  0.784155 \\ 
  $\gamma=1$ & 0.733333 \\
 $\gamma=2$ &  0.68325 \\ 
 $\gamma=3$ &   0.660393 \\
$\gamma=10$ & 0.633238 \\
$\gamma=100$ & 0.627494 \\
\hline
\end{tabular}
\end{center}
\caption{Probability that both eigenvalues are real for a $2 \times 2$
  matrix with matrix elements chosen independently from the symmetric
  Gamma distribution in Eq.~(\ref{eq:gammadef}). The probability lies
  in the interval $[5/8,7/8]$ for $\gamma \in (0, \infty)$, with the lower bound corresponding to $\gamma=\infty$ and upper to $\gamma=0$.}

\label{ta:gamma}
\end{table}

\subsubsection{Power law distributions}

A qualitatively different process is interesting to consider, and as
in the case of random polynomials \cite{LoganShepp1966}, it will be
interesting to study what happens when the underlying probability
distributions have diverging moments. The Cauchy distribution which is the 
simplest and best studied of these and occurs in many contexts, is
given by   
\be
p(x) = \frac{1}{\pi(1+x^2)}
\label{eq:cauchydefn}
\ee
for $x\in(-\infty,\infty)$. Again the possible parameter in the
distribution is rendered inoperative via scaling.  
Using ${\tilde p}(\omega) =
e^{-|\omega|}$ and performing a change of variables
$x = u^2$ in (\ref{eq:Prob2Charact}), the integral w.r.t. $u$ can be
calculated. 
Then computing the integral w.r.t. $\omega$ using series expansion yields
\begin{equation}
P_{2,2}^{(1)} = 1 - \frac{4}{\pi^2}\int^{\infty}_{0}\frac{1}{1+y^2}\tan^{-1}\left(\frac{1}{1+\sqrt{\frac{2}{y}}}\right)dy=\frac{3}{4}.
\label{eq:Cauchy2}
\end{equation}
Although the integral is written as if it is carried out, it is in
fact a numerical evaluation which is almost certainly
correct. Evaluations using the convolution path lead to interesting
alternative integral forms of $3/4$, but none of them (including the
above) seem to be either in standard tables or calculable using symbolic
mathematical packages. 

In general, one may consider the distribution $p_a(x) \sim 1/(1+x^{2
  a})~,~ a \geq 1$ which 
possesses finite (and nonzero) moments up to order $2 a-2 $
only. Numerical evaluation of Eq.~(\ref{eq:Prob2IntglConv}) gives
$0.7076005$ and 
$0.694185$ for $a=2$ and $3$ respectively, both of which are smaller
than the result obtained above for the Cauchy distribution which is
the slowest decaying power law distribution with finite
mean. We also note that  compared to the
Gaussian case, the eigenvalues are less likely be real for the
power law-distributed matrix elements with $a > 2$, and therefore the probability $P_{2,2}^{(1)}$ is not merely determined by 
the decay behavior of the distribution of the matrix elements. Other
distributions such as $1/(1+|x|^{a})~,~1 < a < 2$ which is slower 
  than Cauchy distribution but not smooth, or power laws corrections
  that  vanish or diverge at the origin to the fat-tailed
  distributions have not been investigated.  

To summarize, the case of a single $2 \times 2$ matrix with elements drawn from various i.i.d. distributions have revealed an interesting phenomenology.
It may seem like the weight of the distributions near the origin
matters and if the values are clustered around zero, the probability of
real eigenvalues increases. This statement must however be qualified:
the normal distribution with however a small variance always have only
$1/ \sqrt{2}$ probability of having real eigenvalues. Thus the
differentiability of the underlying  distributions play an important
role. In the extreme case of delta distributions, we may have $100\%$
eigenvalues real. But given that the distributions be smooth to all
orders, the normal distribution seems to be singled out as the one
with the largest probability of real eigenvalues. Also, the probability
of real eigenvalues seems to be in the range $[5/8,7/8]$ for the
classes of distributions considered here. For the case of the Cauchy
distribution which is smooth but has diverging variance, the probability is large at $3/4$ for real eigenvalues. Thus
this rather simple problem of the probability of real eigenvalues of
random real $2\times 2$ matrices seems to possess a multitude of
interesting features that warrants further study and clarification.


\section{Real eigenvalues of product of matrices}
\label{secproduct}

We now turn to the properties of an $n \times n$
matrix obtained after taking a product of $K$ square 
matrices, and study how the probability that some or all 
of the eigenvalues are real behaves for $K > 1$. It is of interest to
see if the results for 
single $2 \times 2$ matrix described in the last section carry over to
the product of matrices. Thus there is a two-fold generalization: (1)
products of matrices are considered, and (2) their dimensionality 
can be more than two.


\subsection{Asymptotic value and maintenance of hierarchy}

\begin{figure}[t]
\includegraphics[width=1.0\textwidth]{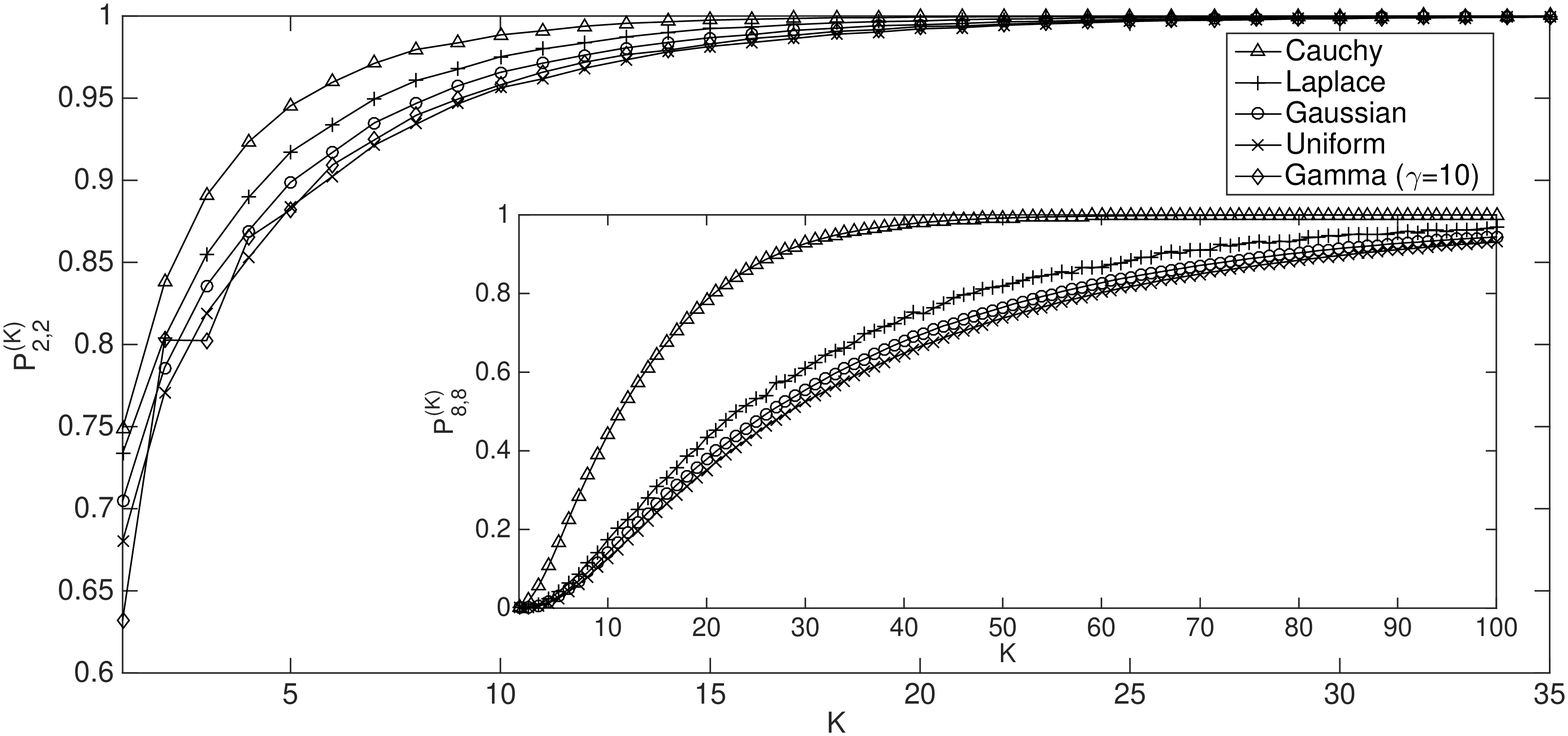}
\caption{Comparison of probability that all eigenvalues are real for
  a product of $K$ random matrices with different
  symmetric distributions and the dimensionality $n=2$ (main) and $8$
  (inset). The plot is based on $10^5$
  independent realizations.}
\label{fig:prodallreal}
\end{figure}

We numerically studied the probability $P_{n,k}^{(K)}$ that $k$ eigenvalues 
are real for a product of $K$ $n \times n$ random matrices. As the 
products of random matrices can have, in 
general, a positive Lyapunov exponent \cite{Furstenberg1960},
numerical procedures for all cases of products ``renormalizes" the
matrices for each product by dividing the Frobenius norm. This however
does not alter the quantities of interest here. For most of the discussion, the matrix 
elements are assumed to be i.i.d. random variables distributed according to one of
the following symmetric probability distributions: uniform, Gaussian,
Laplace and Cauchy. Note that all of these distributions are finite at
the origin, but have different tail behavior. 
The data were averaged over
$10^5-10^6$ independent realisations of the random matrices. For the
$2 \times  2$ matrix, there can be either zero or two real
eigenvalues, but for higher 
dimensional matrix with $n=8$ that we consider here, the
probability of real eigenvalues is nonzero for $k=0,2,4,6$ and $8$
only. Numerical results for the probability of all real eigenvalues for $n=2$ and $8$ are presented in Fig.~(\ref{fig:prodallreal})
for various probability distributions as a function of the number
of matrices in the product. We find that the probability of all 
eigenvalues being real increases to unity with $K$ 
monotonically for most distributions (see, however, the case
of Gamma distribution defined by (\ref{eq:gammadef}) with
$\gamma=10$). This effect has been previously observed by one of the 
authors \cite{Arul2013} for Gaussian-distributed matrix elements. Here we
find that this result is quite general in that the eigenvalues tend to become
real with increasing $K$ when matrix elements are
distributed according to non-normal distributions as well.

\begin{figure}[t]
\includegraphics[width=1.0\textwidth]{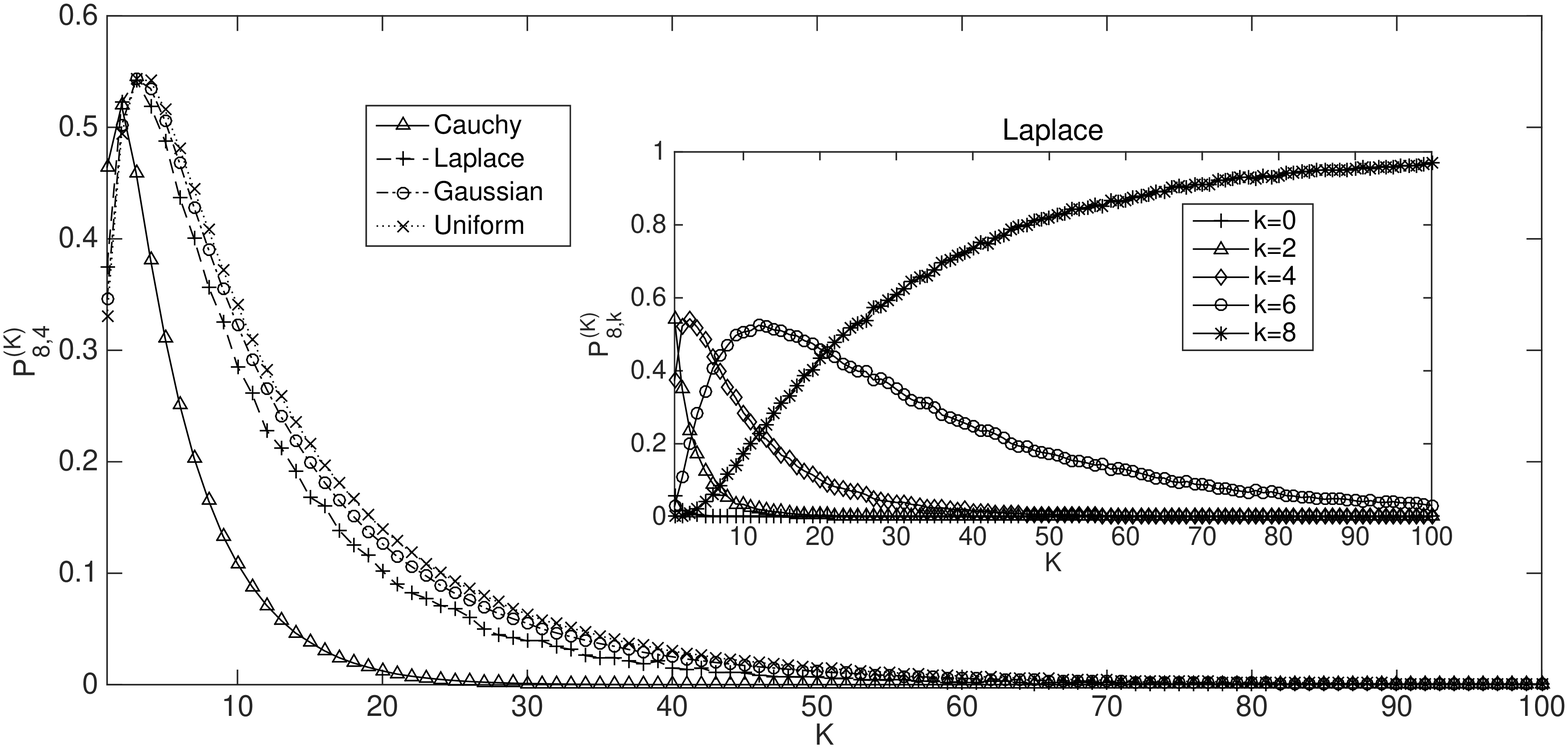}
\caption{Probability that $4$ eigenvalues are real for a product of
  $K$ $8 \times 8$ random matrices whose elements are chosen from
  uniform, Gaussian, 
  Laplace and Cauchy distributions, each with zero mean, based on $10^{6}$
  realizations. The inset shows the probability that $k$ eigenvalues are real for a product of
  $K$ $8 \times 8$ zero mean, Laplace-distributed random matrices.} 
\label{fig:prodallreal_N8}
\end{figure}

The second important point illustrated by Fig.~(\ref{fig:prodallreal})
is that the same hierarchy 
  as observed for the $K=1$ case continues to hold for $K > 1$. 
 Explicitly, the probability of all eigenvalues being real for the
 uniform distribution is the smallest followed by the Gaussian 
 distribution, the Laplace distribution and finally the Cauchy distribution.  
  Thus the slowly decaying distributions appear to have larger
probability of having all real eigenvalues. This feature is seen to be
valid for higher dimensions as well (results not presented). Thus the
hierarchy apparent even with a single $2 \times 2$ matrix continues to
hold for larger dimensional matrices as well as the product of the
random matrices. Furthermore, Fig.~(\ref{fig:prodallreal_N8}) shows that
the ordering of the probability of real eigenvalues according to their
tail behavior is not special to the probability of all real 
eigenvalues, but  holds when $k \neq n$ as well. Note, however, while
the  probability of all real 
eigenvalues increases monotonically towards unity, the 
probability of $k < n$ real eigenvalues decays to zero, as the number 
of matrices in the product increases (see, the inset of
Fig.~(\ref{fig:prodallreal_N8})).

The increase in the probability of all real eigenvalues is 
reflected in the average number of real eigenvalues as well which is
given by $E_n^{(K)} = \sum^{n}_{k=0}k P_{n,k}^{(K)}$. 
Since, as discussed above, the probability $P_{n,k}^{(K)} \to
\delta_{n,k}$ as $K$ increases, the average $E_n^{(K)}$ approaches the
dimension of the matrix. Numerical results in  
Fig.~(\ref{fig:avgexpN8}) for $8 \times 8$ matrices strongly suggest an exponential approach of
the  expected number of real eigenvalues to the dimension of the
matrix. 
The data also points to the continuance of the hierarchy even for the
average number of real eigenvalues indicating a more microscopic
adoption of the hierarchy to all $P^{(K)}_{n,k}$.  

\subsection{Effect of correlations between matrix elements}

\subsubsection{Hadamard product}

We now consider  the Hadamard (or Schur) product wherein the elements
of the  product are
simply the products of the corresponding elements. Thus if $\circ$ is
used to denote the Hadamard product of two matrices $A$ and $B$, then  $(A \circ B)_{ij}=A_{ij}
B_{ij}$.  The Hadamard product of two positive semidefinite matrices
is also positive semidefinite, unlike the usual product, and is one of
the reasons it is important \cite{Hornmatrixanalysis}. However the
reason why we choose to study Hadamard products is 
 that although the matrix elements of the product matrix are products
 of random numbers, there is no correlation amongst the matrix elements.  
Thus one can hope to disentangle two possible mechanisms that may be 
responsible for the phenomenon that all the eigenvalues of a 
random product matrix tend to be real: the act of simple multiplication which
is presumably causing the matrix elements to have large weight near
the origin and the various addition of such products that are leading
to correlations between the matrix elements. 

\begin{figure}[t]
\includegraphics[width=1.0\textwidth]{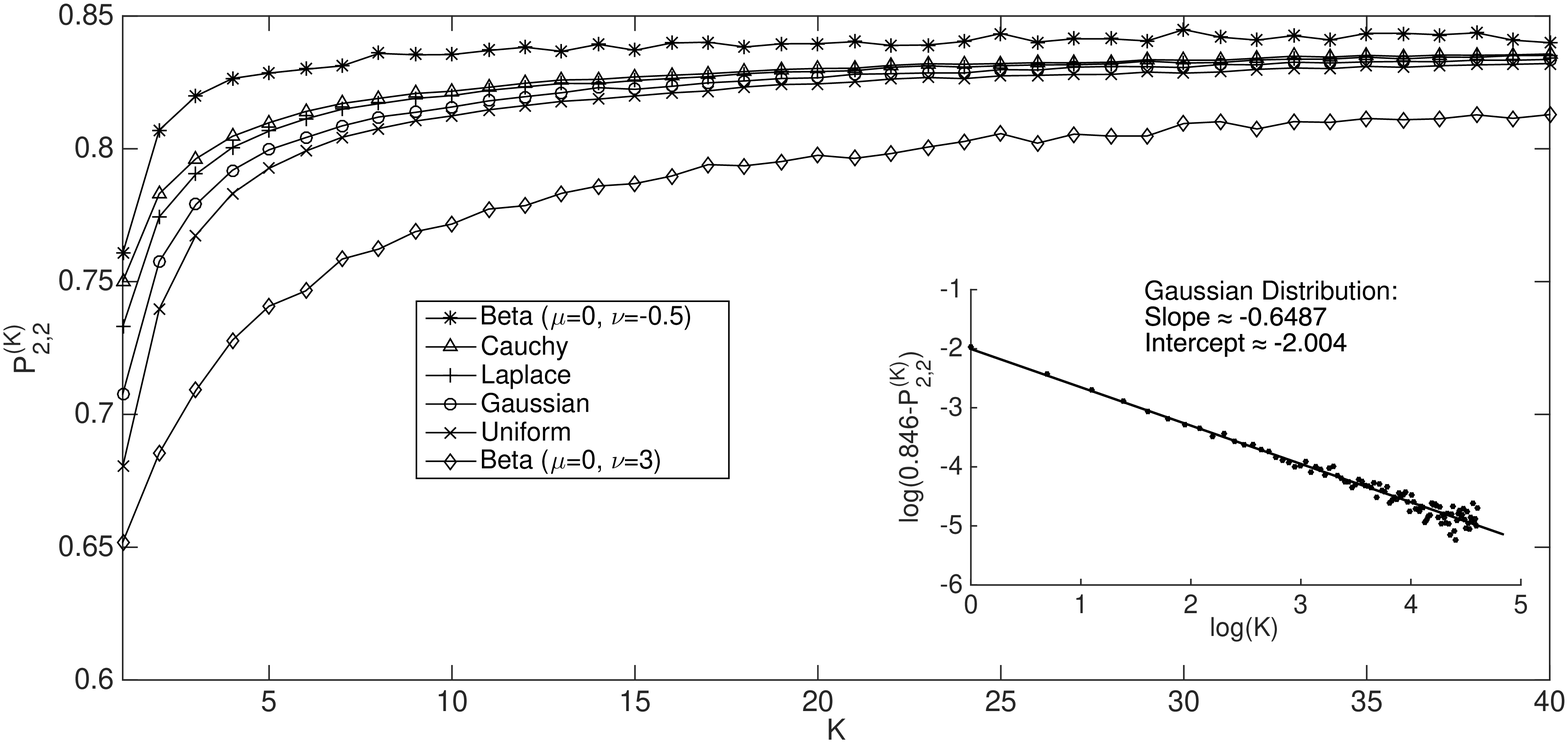}
\caption{Comparison of probability that all eigenvalues are real for
  {\it Hadamard} products of $K $  $2\times 2$ random matrices for some symmetric distributions based on
  $10^5$ realizations. For the Beta distribution with $\nu=3$, the
  x-axis is scaled up by a factor $5$ since the convergence to the
  asymptotic value occurs for very large $K$. The inset shows the
  power law approach of the 
  probability of all real eigenvalues to the asymptotic value which is
  less than unity, for the Gaussian case. The plot is based on
  $10^5$ realizations for the Beta distribution, and $10^6$ for the rest.}
\label{fig:hadamard}
\end{figure}

To this end, we consider the eigenvalue properties of a Hadamard
product of $K$ random $n\times n$ matrices with elements distributed
according to uniform, Laplace, Gaussian and Cauchy distributions, each
with zero mean. Figure~(\ref{fig:hadamard}) shows the case of Hadamard
products of $2\times 2$ matrices, and we observe that the hierarchy
effect is maintained at any $K$ (although there are more fluctuations in
this case as compared to that of the ordinary matrix product). But,
importantly, we find that the probability does {\it not} approach
unity with increasing $K$.

We can get some insight into this result by calculating the distribution
of the matrix elements of the Hadamard product matrix for some
distributions, and appealing to the
results for the $2 \times 2$ matrices obtained earlier in Sec.~\ref{single}.  
Let us first consider the case of Hadamard product of matrices with elements drawn from the bounded distribution $p_{0,\nu}(x)$ defined in Eq.~(\ref{bddef}). Let $z_K=x_1x_2...x_K$ represent a random variable formed by the
product of $K$ i.i.d. random variables. As discussed in Appendix~\ref{appC}, the distribution of the product for this case is given by 
\be
\begin{split} 
p_K(z_K=x_1x_2...x_K)=\frac{(\nu+1)^K |z_K|^\nu }{2 (K-1)!} ~ \left[\ln\left(\frac{1}{|z_K|}\right)\right]^{K-1}  \Theta(1-|z_K|). 
\label{unifhadK}
\end{split}
\ee
For $\nu \leq 0$, the above distribution diverges at the origin while for positive $\nu$, it vanishes at $|z|=0$ and $1$ and is a nonmonotonic function of $|z|$. Thus, except for the uniform case, the behavior of the distribution $p_K(z_K)$ is similar to that of $p_{0,\nu}$ near the origin. 
Using the above result, it is possible to numerically evaluate the integral in (\ref{eq:Prob2IntglConv}) to obtain $P^{(K)}_{2,2}$ for various values of $\nu$ and $K$. For example, for $\nu=0$, we find the probability of real eigenvalues to be $0.738779$, $0.767331$, $0.782558$, $0.792032$, $0.798561$, $0.803376$ for $K = 2,3,4,5,6,7$ respectively. 
For larger $K$, numerical integration does not converge; however, as
shown in Fig.~\ref{fig:hadamard} for some representative values of
$\nu$, the data obtained from direct sampling indicates the approach  
to a probability less than $7/8$. For the case of uniform distribution, the
probability seems to saturate around $0.84$. But for negative and positive $\nu$, the
probability of real eigenvalues approaches a value higher and lower than $0.84$ respectively. This pattern is consistent 
with the results in Sec.~\ref{single} where the probability of real eigenvalues decreased with increasing $\nu$.

\begin{figure}[t]
\includegraphics[width=1.0\textwidth]{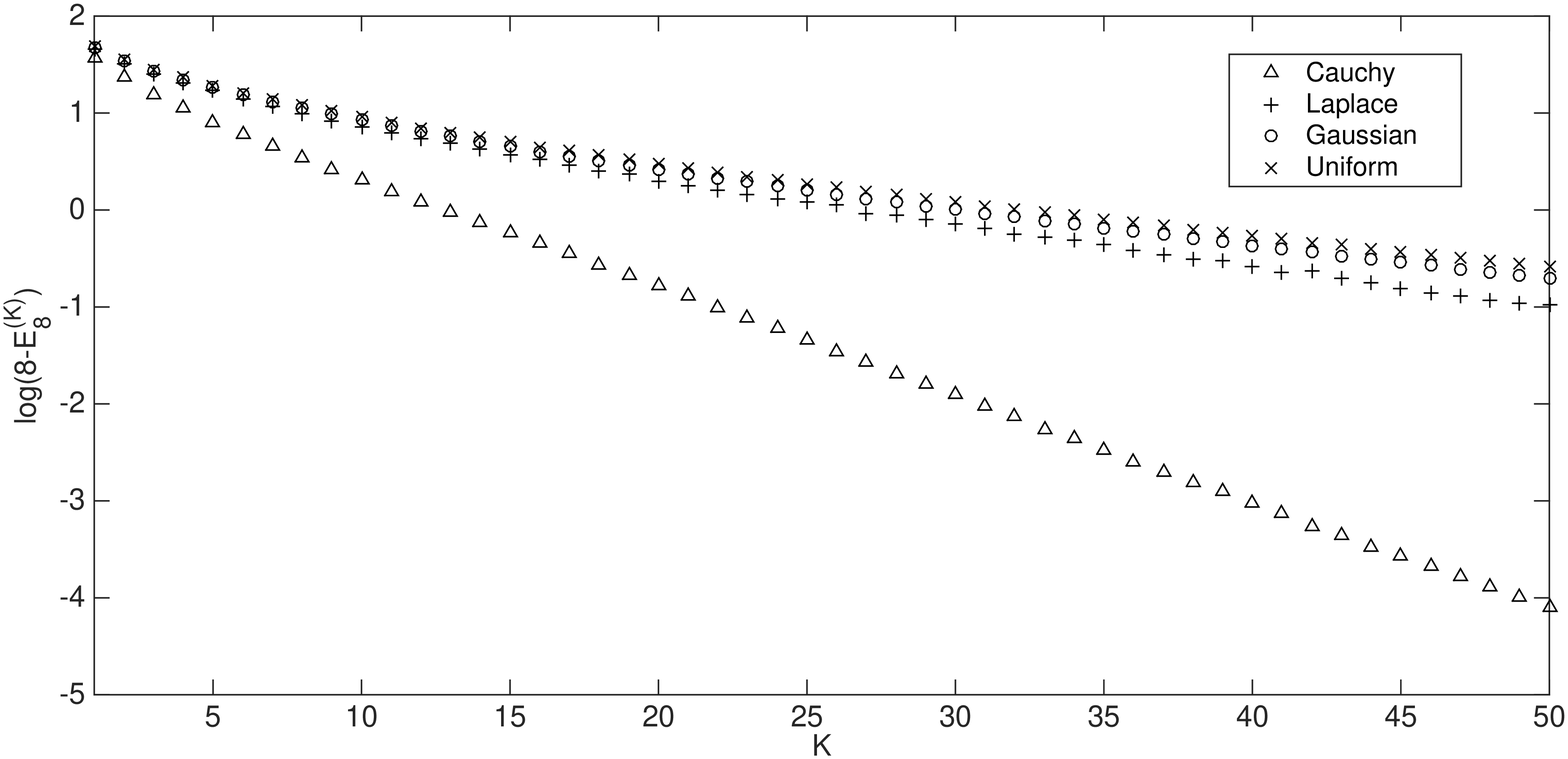}
\caption{Expected number of real eigenvalues for a product of $K$ random $8 \times 8$ matrices with elements chosen from uniform, Gaussian, Laplace and Cauchy distributions, each with zero mean. The plot is based on $10^{6}$ realizations.}
\label{fig:avgexpN8}
\end{figure}

We next consider unbounded distributions, and start with the case of Hadamard product of matrices with Gaussian distributed elements with zero mean and unit variance for which \cite{Springer1970,Forrester2014} 
\be
\begin{split}
p_2(z_2)=~\frac{1}{2\pi}\int_{-\infty}^{\infty}
\int_{-\infty}^{\infty}e^{-x_1^2/2}e^{-x_2^2/2}~\delta(z_2-x_1x_2)
~dx_1 dx_2  \\=\frac{1}{2\pi}\int_{-\infty}^{\infty} e^{-|z_2|\cosh
  x_1}~dx_1 = \frac{K_0(|z_2|)}{\pi} ,~z_2 \in (-\infty,\infty).
\label{gausshadK2}
\end{split}
\ee
which diverges logarithmically at the origin since $K_0(z) \to \ln (1/z)$
as $z$ approaches zero. For this distribution, the integral in  Eq.~(\ref{eq:Prob2IntglConv}) can be
numerically evaluated by using the fact that its convolution is given
by a Laplace distribution (shown in Sec. III B 2) yielding
$P^{(2)}_{2,2}=0.757164$, which is consistent with the value obtained
in Fig.~(\ref{fig:hadamard}) by direct sampling. For $K>2$, the
probability distributions are given by the Meijer-G functions
\cite{Springer1970,Forrester2014} making it difficult to handle this
using  Eq.~(\ref{eq:Prob2IntglConv}). We therefore give the results
obtained  via numerical sampling in Fig.~\ref{fig:hadamard} which again indicates an approach towards $0.84$. 
In the case of Hadamard product of two matrices with Laplace-distributed elements with zero mean given by (\ref{eq:gammadef}) with $\gamma=1$, we have \cite{Springer1970}
\be
\begin{split}
p_2(z_2)=~\frac{1}{4}\int_{-\infty}^{\infty} \int_{-\infty}^{\infty}e^{-|x_1|}e^{-|x_2|}~\delta(z_2-x_1x_2) ~dx_1 dx_2 \\=\frac{1}{2}\int_{-\infty}^{\infty} e^{-2\sqrt{|z_2|}\cosh x_1}~dx_1 = K_0(2\sqrt{|z_2|}),~z_2 \in (-\infty,\infty),
\label{laphadK2}
\end{split}
\ee
where the last step in Eq.~(\ref{laphadK2}) has been done by a series of two transformations, $x_1\rightarrow x_1\sqrt{|z_2|}$ followed by $x_1\rightarrow e^{x_1}$. 
Numerical evaluation of Eq.~(\ref{eq:Prob2IntglConv}) using this yields
$P^{(2)}_{2,2}=0.773849$, and the results obtained from direct
sampling show a convergence towards a value close to $0.84$. For the
Cauchy-distributed matrix elements, the distribution of the product of
two random variables is known to be \cite{Springer1966}
\be
p_2(z_2)=\frac{\ln z_2^2}{\pi^2 (z_2^2-1)},
\ee
which diverges at the origin logarithmically.

Thus, as Fig.~(\ref{fig:hadamard}) shows, the probability of all real 
eigenvalues monotonically increases with $K$ and saturates to 
a value in the range $0.82-0.84$. Note that these numbers are less than
$7/8$ which, from our analysis in the previous section, is expected
of  distributions that diverge as a power law, $|x|^{\nu}$ with $\nu
\rightarrow  -1$, at the origin. 
From the above calculations for $K=2$, it is clear that the
uniform distribution and the three unbounded
distributions discussed above have Hadamard product elements
distributed according to a probability distribution that diverges as
$\ln(1/|x|)$ at the origin. In fact, the elements of the Hadamard
product of uniform distribution as well as Cauchy distributed random
matrices have probability distributions that diverge as
$\ln(1/|x|)^{K-1}$  at arbitrary $K$ \cite{Springer1966}. It is
reasonable to expect that the Gaussian and Laplace cases also diverge
in a similar manner. For these cases, as the divergence at the origin
is somewhat ``slower" than a power law, it seems reasonable that the probability of eigenvalues being   
real is a shade smaller than that of single $2 \times 2$
matrices whose elements are distributed according to the power law
discussed above. 

Assuming that  the probability of Hadamard products saturate to the
same value for these four distributions in the $K \rightarrow \infty$ limit, and taking  this to be $\approx 0.846$ (in the case of $n=2$), our numerical results show a power law approach to the constant:
\begin{equation}
P^{(K)}_{2,2}=P^{(\infty)}_{2,2} -\dfrac{C}{K^{\theta}},
\end{equation}
where $C$ is a positive constant and the exponent $\theta \approx 0.675, 0.649,
0.654, 0.621$ for uniform, Gaussian, Laplace and Cauchy distribution
respectively (see the inset of Fig.~(\ref{fig:hadamard})). 
Thus, here the probability of real eigenvalues approaches the
asymptotic value as a power law in comparison to the ordinary matrix
product case, where this approach is seen to be exponentially fast
\cite{Arul2013}. The exponential behavior is also seen in the expected
values of real eigenvalues as in Fig.~\ref{fig:avgexpN8}. 
We also looked at the fraction of expected number of real eigenvalues
for higher dimensional matrices ($n=2,3,4$) when each matrix in the Hadamard product has Gaussian-distributed
elements with zero mean, and find that the fraction of real eigenvalues increases and saturates to a value less than one, unlike that
observed for the case of usual matrix products. We also find that the asymptotic value
of this fraction decreases with an increase in 
the dimensionality of the matrices. 

The convergence of the probability that all eigenvalues are real as $K
\to \infty$ for the Hadamard product can also be understood using the
Central Limit Theorem applied to the logarithm of the absolute value
of the product random variable. The elements of the Hadamard product matrix would
then admit a symmetrized log-normal probability distribution of the
form \cite{Sornette:2000}
\be
\begin{split} 
p_K(z_K=x_1x_2...x_K)=\frac{1 }{2|z_K|\sqrt{2\pi K \sigma^2}} ~\exp \left[ -\frac{(\log|z_K|-K\mu)^2}{2K\sigma^2} \right],~z_K \in (-\infty,\infty),
\label{eq:logn}
\end{split}
\ee
for large $K$, with the parameters $\mu$ and $\sigma$ being the mean
and standard deviation of $\log|x|$. Numerically, we find that the
probability that all eigenvalues are real for a matrix with elements
distributed according to Eq.~(\ref{eq:logn}) indeed converges to a
value around $0.84$ at large $K$, as for the Hadamard product.

\subsubsection{Usual matrix product}

We now return to the properties of the matrix obtained after
multiplying $K$ matrices, and ask why the asymptotic probability of
real eigenvalues saturates to unity unlike that for the Hadamard product. We can think of two possible reasons: one, the (usual) product matrix elements have probability distributions
that are significantly different from the Hadamard case. The other possibility is that the
 correlations between the matrix elements in the case of usual products lead
  to an increase in probability of all eigenvalues real to unity as
 opposed to the Hadamard case where this is not true. It is not obvious whether the difference in the probability distribution or the correlations between the matrix elements is responsible for a higher $P^{(K)}_{2,2}$. 

\begin{figure}[t]
\includegraphics[width=1.0\textwidth]{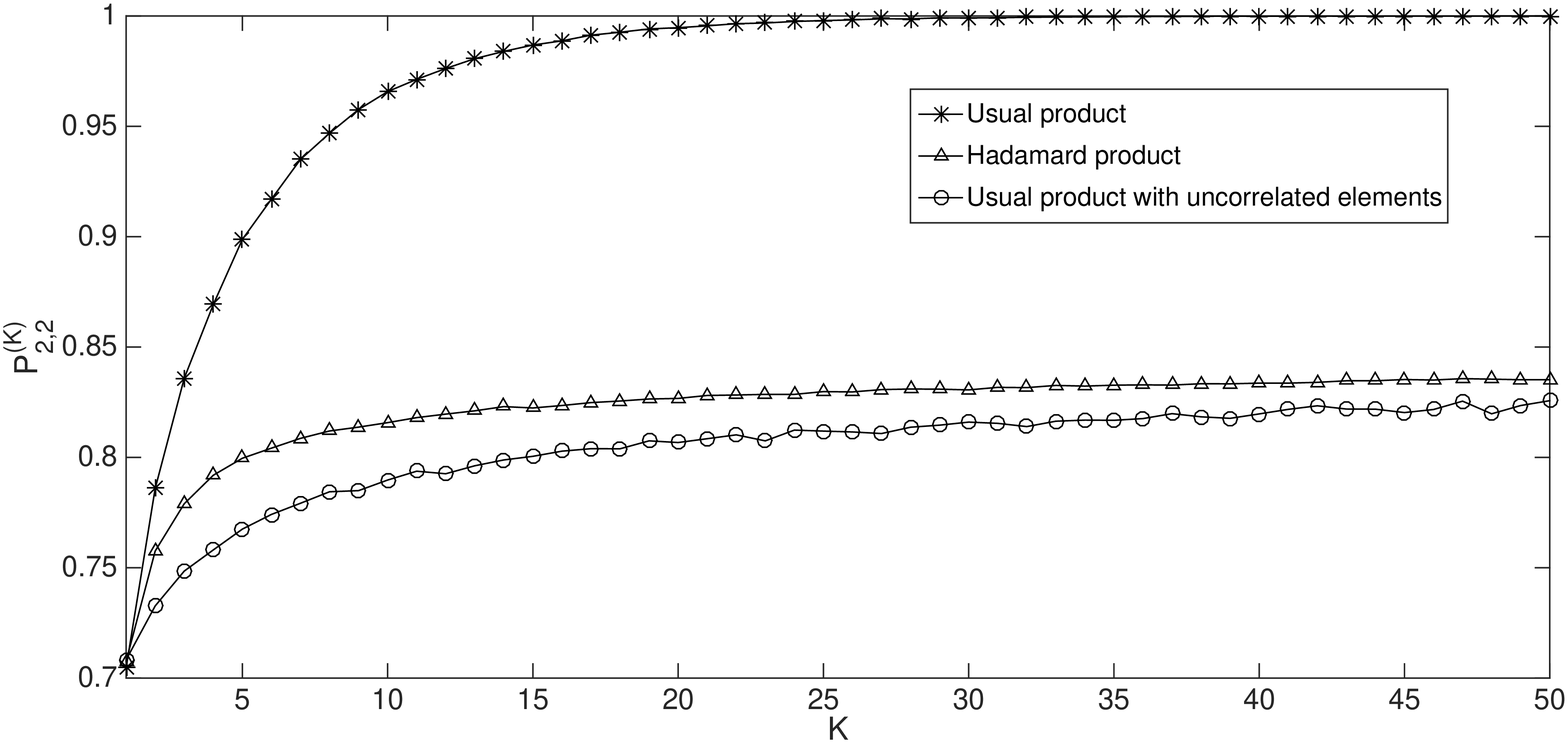}
\caption{Comparison of $P^{(K)}_{2,2}$ for Hadamard products and
  usual products with correlated elements and $P^{(1)}_{2,2}$ for
  matrices with i.i.d elements distributed according to  probability
  distribution of product of $K$ independent matrices, for $2\times 2$
  zero mean, Gaussian-distributed random matrices. The plot is based on
  $10^6$ realizations for the Hadamard product, and $10^5$ for the rest.}
\label{fig:ProdDist}
\end{figure}

To gain some insight into this intriguing question, we  looked at the matrices with i.i.d. random 
variables as elements, each distributed according to the probability 
distribution of the product matrix elements at various $K$. Such matrices 
can be easily generated by taking independent samples of 
usual product matrices at each $K$ and forming new matrices by 
picking the corresponding matrix elements from the independent 
product matrix samples. These new matrices would then have elements 
with probability distributions corresponding to that of elements of a 
usual product matrix at product length $K$, but the matrix elements 
would now be independent, having been selected from independent 
product matrix samples. Fig. (\ref{fig:ProdDist})
shows a comparison between $P^{(K)}_{2,2}$ for the 
Hadamard product and usual product and $P^{(1)}_{2,2}$ for
matrices with probability distribution of elements corresponding 
to that of usual product at product length $K$, but without any 
correlations between the elements. It is clear that the difference
 in probability distribution of matrix elements between the Hadamard and the 
 usual product has an effect of lowering $P^{(K)}_{2,2}$ for 
 the usual product case to a value below that of the Hadamard product case. 
 However, the presence of correlations between 
 the matrix elements of the usual product matrix is seen to have an effect 
 of a substantial increase in $P^{(K)}_{2,2}$, finally leading to 
 an asymptotic value of one.

The effect of correlations can be seen analytically for the case of Gaussian-distributed matrices when $K=2$. We have seen 
 that the distribution of the product of two Gaussian-distributed random 
 variables with zero mean and unit variance is given by Eq.~(\ref{gausshadK2}) and the probability of real eigenvalues is  $0.757164$.  
 The distribution of the matrix elements obtained on taking the usual matrix product corresponds 
 to that of the sum of two random variables, each distributed 
 according to the distribution of the product of two random variables. 
The characteristic function of Eq.~(\ref{gausshadK2}) is given by ${\tilde p}(k)=(1+k^2)^{-1/2},~k\in[-1,1]$ (due to Eq.~(11.4.14) of \cite{Abramowitz1964}). 
It is easily 
seen that this ${\tilde p}(k)$ is the square root of the characteristic 
function of Laplace distribution. Hence, the product matrix 
elements have a Laplace distribution, for which we know from 
Eq.~(\ref{eq:laplace}) that $P^{(1)}_{2,2}$ would have been $11/15 = 0.7333$, 
had the elements been independent. 
But, the correlations between the matrix elements raise this probability 
to $\pi/4 = 0.7854$ \cite{Arul2013,Forrester2014} for the usual product, hence 
presenting a strong evidence that the correlations between the 
matrix elements are in fact leading to $P^{(K)}_{2,2}\rightarrow 1$ 
observed for usual products.

We also measured the correlations between the matrix elements to see how correlations increase with increasing $K$ for $2
  \times 2$ matrices with Gaussian-distributed matrix elements. More 
  precisely, we consider
\be
C_i= \left(\frac{\langle x_1 x_i \rangle-\langle x_1 \rangle \langle
  x_i \rangle}{K} \right)^{1/K}   ~,~i=1,...4
\ee
where $x_i$ are the matrix elements obtained after taking the product
of $K$ matrices. For $K=1$, obviously $C_i= \delta_{i,1}$. But with increasing
$K$, all the four $C_i$'s seem to be of same order {\it i.e.} the
variance of the new distribution and the correlations are
similar. In particular, we find that for $K=2$, the correlations
$C_i=1.417, 0.047, 0.031, 0.057$ for $i=1,2,3,4$ respectively, but
these numbers increase to $1.885, 1.446, 1.349, 1.462$ and
$1.905, 1.500, 1.475, 1.667$ for $K=12$ and $20$
respectively. These data suggest that similar to the probability
$P_{2,2}^{(K)}$, the correlations also tend to saturate with
increasing $K$.

\section{Discussion}

How does the probability of real eigenvalues for a $2 \times 2$ random
matrix depend on the probability distribution of the matrix? 
It is quite surprising that such an apparently elementary question
presents many novel  challenges, and throws up some surprises. Here we
find that the probability of real eigenvalues depends on several
detailed features of the probability distribution, some of which we have
identified here as the finiteness and smoothness at the origin,
existence of maxima away from origin and the finitness of the moments. 
We have shown that eigenvalues are most
likely to be real for distributions limiting to $p(x) \sim 1/|x|$ which is the most divergent distribution  at the
origin (with zero mean) and occur with a probability $7/8=0.875$. That
the large
weight at the origin correspond to high probability is perhaps not
surprising since when the matrix elements are distributed according to 
$\delta(x)$, this probability is unity. However, interestingly, here we 
find the maximum probability to be less than one. The origin of this naturally rests in
the smoothness of the probability distributions considered. We also mention that
the values are not universal; for example, the probability turns out to be
different for the Beta distribution $p_{0,2}(x)$ and for Gamma distributed matrix elements
with $\gamma=3$, although both have the same behavior near $x=0$.

Our results suggest that the probability that all eigenvalues are
real is larger for distributions with large weight at the origin and 
that decay slowly. More concretely, there exists a hierarchy between the 
probability values of the different distributions, which is Cauchy
(0.750) \begin{math}>\end{math} Laplace
  (0.733) \begin{math}>\end{math} Gaussian
    (0.707) \begin{math}>\end{math} uniform
      (0.6805). Moreover, for class of distributions that decay in the same manner, the distributions with
higher weight close to zero have a higher probability of real
eigenvalues. For example, although the probability $p_{2,2}^{(1)}$ for
distribution Eq.~(\ref{eq:gammadef}) for $\gamma \leq 2$ is larger than the
uniform distribution as it decays slower than the bounded
distribution, this probability for $n=3$ is found to be smaller than
that for the uniform case. Thus the `reality' is determined by both
the tail and near-zero behavior of the probability distribution from which the
matrix elements have been chosen.   
For the broad class of smooth  distributions we have considered, we did  not find 
the probability of real eigenvalues to lie outside the range $[5/8,7/8]$, which appear
to be natural boundaries for sufficiently smooth distributions. 
 
Numerical
      results discussed above also shows that this remains valid even
      after taking a product of random matrices. It is quite
      surprising that the curves for probabilities of the
      different distributions do not cross each other. An intuitive
      explanation of this is yet to be figured out.  We also measured
      the distribution of the matrix elements of the product matrix,
      and find that it decays slower (but faster than any power law)
      as the number $K$ in the product increases. This observation is
      consistent with the result for single matrices (with finite
      moments),  namely, that faster the distribution decays, smaller is the probability that all eigenvalues are real. However this is only
part of the explanation as the probability $p_{N,N}^{(K)}$ always
stays below unity when correlations between matrix elements are
ignored, and in order to approach unity, correlations are found to be 
crucial. It maybe stated that despite extensive and exact results
obtained so far there is not much understanding about {\it  why} the
eigenvalues tend to become real under the usual matrix product, and
more work in this direction is clearly desired.

\acknowledgements{AL thanks JNCASR, Bangalore, for supporting a short visit 
facilitating discussions. It is a pleasure to thank an anonymous referee for suggesting the proof that $5/8$ is the lower-bound.}

\newpage


\appendix

%
%


\section{Detailed derivations for symmetric beta distribution}
\label{app_beta}

When $\mu=0$ and $\nu=2 k$ is a positive integer, the convolution can
be written in the form of a finite series: 
\be
q(z)=\Theta(2-|z|)\left( \frac{2k+1}{2}\right)^2 \sum _{r=0}^{2k} \binom{2k}{r} (-|z|)^{2k-r} \dfrac{1-(|z|-1)^{2k+r+1}}{2k+r+1}.
\label{eq:CovModxpownu}
\ee
Integrations can be done using the hyperbolic variables as the combination $xy$ appears exclusively.
To  outline the method used, hyperbolic coordinates ($v,w$) where $x=v e^w$ and $y=v e^{-w}$ are useful. Thus the variable $w$ does not appear in the transformed expression. The range of $v$ is $[0,1]$, while for a given $v$, the range of $w$ is $[\ln v, -\ln v]$. The integration over $w$ is easily carried out first, and noting that the Jacobian of the transformation is $2v$, leads finally to: 
\be
\begin{split} P^{(1)}_{2,2}(\mu=0,\nu=2k)=1-(2k+1)^4 \times
 \sum_{r=0}^{2k} \left[ \binom{2k}{r} \dfrac{2^{2k-r+1}}{2k+r+1} \left( \dfrac{(-1)^r}{(2k-r+1)(6k-r+3)^2} \right. \right.+\\
\left. \left. \sum_{l=0}^{2k+r+1} \binom{2k+r+1}{l} \dfrac{(-2)^{l}}{(2k-r+l+1)(6k-r+l+3)^2} \right) \right].
 \end{split}
\label{eq:Probmodxpownu}
\ee
As special cases:
\be
P^{(1)}_{2,2}(\mu=0,\nu=2)=\dfrac{8905}{14112} = 0.631023, \; P^{(1)}_{2,2}(\mu=0,\nu=4)= \dfrac{45332489}{72144072} \approx 0.628361.
\ee
The case of odd integer $\nu$ is also accessible, however we only
state the result when $\nu=1$. The convolution is now piecewise
continuous on $|z|<1$ and $1<|z|<2$, 
and the probability of real eigenvalues is
\be
P_{2,2}^{(1)} (\mu=0,\nu=1)=\frac{3653}{5760}+\frac{\ln 2}{240} \approx 0.63709.
\ee

To understand the behavior of the probability of real eigenvalues for 
negative $\nu$, it is convenient to write the convolution for the
distribution $p_{0,\nu}(x)$ defined by Eq.~(\ref{bddef}) as
\beq
\begin{split}
&q(z)=\int_{-\infty}^{+\infty}p(x)p(z-x) \,dx \\& = \left( \frac{\nu+1}{2 } \right)^2 \int_{-1}^{+1} |x|^{\nu} |z-x|^{\nu}  \Theta(1-|z-x|)\, dx.
\end{split}
\eeq
Note that $|x|\le 1$ and $|z-x|\le 1$ implies that $|z|\le 2$. From the symmetry $q(z)=q(-z)$ it suffices to consider $0\le z \le 2$.
Using the hyperbolic coordinates, $x=v e^w$ and $y=v e^{-w}$, we get 
\beq
\begin{split}
& P_{2,2}^{(1)} =1-(\nu+1)^2 \int_0^{1} 2v \, dv v^{2 \nu}  \int_{\ln v}^{-\ln v } dw \int_0^{2 v} dz  q(z)\\ &=1+(1+\nu)^4 \int_{0}^1 v^{1+2 \nu} \ln v \int_0^{2 v} dz\, \int_{-1}^1 |x|^{\nu} |x-z|^{\nu} \Theta(1-|x-z|) dx\, dv.
\end{split}
\label{eq:nunegmainApp}
\eeq
Except when $\nu$ is an even integer, some care must be taken 
as $q(z)$ is piecewise continuous in the two intervals $[0,1]$ and $[1,2]$. 
For $\nu=-1/2$, we have 
\be
P_{2,2}^{(1)} (\mu=0,\nu=-1/2)=\frac{1}{48}(41-\pi -2 \ln 2) \approx 0.759836.
\label{Probnuminhalf}
\ee

To find the probability as $\nu \to -1$, consider the two inner
integrals over $z$ and $x$ in
Eq.~(\ref{eq:nunegmainApp}). On considering various cases arising due
to the Heaviside theta function and performing the integral over $z$
first, we get   
\be
P_{2,2}^{(1)} (\mu=0,\nu) = \frac{3}{4}-\frac{\Gamma(\frac{1}{2})
  \Gamma(2 + \nu)}{\Gamma(\frac{3}{2} + \nu)}~\frac{1 + 4 (1 + \nu)
  \ln 2}{4^{2 (2 + \nu)}} -J_1+J_2, 
  \label{J1J2}
\ee
where 
\be
J_1=(1+\nu)^2 \left(\int_0^{1/2} dv~ v^{1+2 \nu} \ln v~ (1-2 v)^{1 + \nu} -\int_{1/2}^1 dv~ v^{1+2 \nu} \ln v ~(2 v-1)^{1 +
  \nu} \right), 
\ee
and
\be
J_2= 
(1+\nu)^3 \left( 2 \int_0^{1/2} dv ~v^{2 + 2 \nu} \ln v
\int_0^{1-2v} dx~ (x (2 v + x))^\nu + \int_{1/2}^1 dv ~v^{1 + 2 \nu} \ln v
\int_{2v-1}^1 dx ~x^\nu (2 v - x)^{1+\nu} \right).
\ee
It is clear that when $\nu \to -1$, the second term on the right hand side of both 
$J_1$ and $J_2$ is zero. The first integral in $J_1$ contributes
$-1/4$ when the integral is carried out by expanding the integrand
around $v=0$.  However, the dominant contribution to the first
integral in $J_2$ comes when both $x$ and $v$ are small. Then it is
useful to split the integral over $v$ from $0$ to $1/4$ and $1/4$ to
$1/2$ and it turns out that the former  integral with $x$ lying in the
range $0$ to $2 v$ contributes $-1/16$ to $J_2$. Using these results and
setting $\nu=-1$ in the second term in Eq.~(\ref{J1J2}), we finally
obtain the desired result, viz., $P_{2,2}^{(1)} (\mu=0,\nu \to
-1)=7/8$.  


\section{Detailed derivations for smooth bounded distributions}
\label{app_smooth}

For the case $\eta=1$ (parabolic distribution), the convolution is still calculable as: 
\be
q(z)=\frac{3}{160}(2-|z|)^3 (4+6 |z|+z^2) \Theta(2-|z|).
\ee
The integrals in Eq.~(\ref{eq:Prob2IntglConv}) are then elementary and are easily done with mathematical packages, and yield the probability
\be
P_{2,2}^{(1)}(\eta=1)= \dfrac{489341}{705600} = 0.69351.
\ee
The case for $\eta=2$ leads to longer expressions, but with
Mathematica, it is easy to evaluate the convolution and the resultant
probability as  
\be
q(z)=\dfrac{5}{14}\dfrac{1}{16^2} (2-|z|)^5(16+40|z|+36 z^2+10 |z|^3+z^4)\Theta(2-|z|),
\ee
and 
\be
P_{2,2}^{(1)}(\eta=2)= \dfrac{180521487191}{258564354048} \approx 0.698168.
\ee
Similar analysis yields the probability for $\eta=5, 10, 20, 50$ as
  $0.702769, 0.704785, 0.705906, 0.706616$ respectively.  
A monotonic convergence of the probability to $1/\sqrt{2}$ as $\eta$
increases is then a reasonable indication from these calculations. 

\section{Detailed derivations for Gaussian distribution}
\label{app_gauss}

For the Gaussian distribution (\ref{eq:Normal}), the convolution with itself is also a Gaussian distribution but with variance of $2$ rather than $1$: $q(z) = e^{-z^2/4}/(2 \sqrt{\pi})$.
Application of Eq.~(\ref{eq:Prob2IntglConv}), along with the usage of hyperbolic coordinates results in 
\bea
P^{(1)}_{2,2} &=& 1-\frac{1}{\pi^{3/2}} \int_{0}^{\infty}
e^{(x^2+y^2)/2} \int_0^{2 \sqrt{x y}} e^{-z^2/4} dz\; dx\, dy \\
&=& 1-\frac{1}{\pi^{3/2}} \int_0^{\infty}2 v dv \int_{-\infty}^{\infty} e^{-v^2 \cosh 2 w} dw \int_0^{2 v} e^{-z^2/4} dz.
\eea
The integral over $v$ becomes elementary when the burden of a finite range of integration is shifted from the $z$ variable to the variable $v^2$. That is, the $v$ integral is done first following writing the $z$ integral as $e^{-z^2/4} \Theta(-z +2 v)$ over the interval $[0, \infty)$. The probability of real eigenvalues  simplifies to 
\be
P^{(1)}_{2,2}=1-\frac{2}{\pi \sqrt{2}}\int_0^{\infty} \dfrac{dw}{\cosh 2 w \cosh w}= \dfrac{1}{\sqrt{2}}.
\ee
This follows as the integral can be evaluated as $\pi(\sqrt{2}-1)/2$ either with packages such as  Mathematica, or with the use of the residue theorem. This is applied to a rectangular contour with base on the interval $[-L,L]$, and the top on the interval with $\mbox{Im}(w)=i \pi$ and $\mbox{Re}(w)\in [-L,L]$. This region encloses three poles of order $1$ at $i\pi/4, \, i \pi/2$ and $i 3 \pi/4$.  In the limit of large $L$, the contour integral is twice what is required.

If we follow the path of the characteristic function, we have 
${\tilde p}(\omega) = e^{-\omega^2}.$ 
Computing the integral in Eq.~(\ref{eq:Prob2Charact}) yields:
\begin{equation}
P_{2,2}^{(1)} = 1 - \frac{2}{\pi\sqrt{\pi}}\sum^{\infty}_{n=0}\frac{(-1)^n}{n!(2n+1)}2^{n-\frac{1}{2}}\left[\Gamma\left(\frac{n}{2}+\frac{3}{4}\right)\right]^2.
\label{eq:SeriesGaussian1}
\end{equation}
A numerical approximation to Eq.~(\ref{eq:SeriesGaussian1}) using
Mathematica yields a value which is exactly equal to
$1/\sqrt{2}$. Also, using the properties of Gamma functions, it is
possible to modify this series to obtain $$P_{2,2}^{(1)} = 1 -
\frac{1}{2\pi}\sum^{\infty}_{n=0}\frac{(-1)^n}{n!}\frac{\Gamma\left(n+\frac{1}{2}\right)\Gamma\left(\frac{n}{2}+\frac{3}{4}\right)}{\Gamma\left(\frac{n}{2}+\frac{5}{4}\right)},$$
which has already been shown to be exactly equal to
$1/\sqrt{2}$ in \cite{Arul2013}.  

\section{Detailed derivations for Gamma distribution}
\label{app_gamma}

For $\gamma=1$, we have the so-called Laplace distribution
$p(x)=(1/2) e^{-|x|}$ for which $p_\gamma(0)$ is nonzero. Using the
linear transformation formula (15.3.3) of \cite{Abramowitz1964} and the 
definition of the Gauss  hypergeometric function ${_2}F_1(a,b;c;z)$ in Eq.~(\ref{eq:Prob2Charact}), after some algebra, we
obtain 
\bea
P_{2,2}^{(1)}(\gamma=1) &=& 1 - \frac{1}{2}\int^{\infty}_{0} d \omega (1 +
\omega^2)^{-7/2} \\
&=& \frac{11}{15} \approx 0.733333.
\label{eq:laplace}
\eea
For integer $\gamma$, similar steps can be used to find an analytical
expression for $P_{2,2}^{(1)}$.

The convolution is given by 
\begin{equation}
q(z)=\dfrac{1}{4 \Gamma(\gamma)^2} \int_{-\infty}^{\infty} |x-z|^{\gamma-1} |x|^{\gamma-1} e^{-|x|} e^{-|x-z|} dx.
\label{eq:Convgamma1}
\end{equation}
Note that the general symmetry $q(z)=q(-z)$ holds, and we can therefore restrict attention to $z\ge 0$. This simplifies the computation somewhat and leads to 
 Thus the convolution is itself the sum of a Gamma
distribution and a Bessel distribution \cite{McKay1932}. If the
distribution was not symmetrized, that is the distribution were the
usual Gamma distribution on $[0, \infty)$, only a term proportional to
  the first will be in the convolution. The second term represents the
  convolution of the distributions on the opposite sides of the
  maxima. 
Now to evaluate the probability of real eigenvalues of a $2\times 2$
matrix with entries from the symmetric Gamma distribution, we use the
first equality in Eq.~(\ref{eq:Prob2IntglConv}), and obtain  
\begin{equation}
P_{2,2}^{(1)}(\gamma) =\dfrac{1}{2}+\frac{1}{\Gamma(\gamma)^2} \int_0^{\infty} dx \int_0^{\infty} dy\, (xy)^{\gamma-1} e^{-x-y} \int_{2 \sqrt{xy}}^{\infty} dz \,q(z). 
\end{equation}
Once again, using the hyperbolic coordinates, $x=v e^{w}/2, y= v e^{-w}/2$ and performing the integral over $w$ results in 
\begin{equation}
P_{2,2}^{(1)}(\gamma) =\dfrac{1}{2}+\frac{1}{2^{2
    \gamma-2}\Gamma(\gamma)^2} \int_0^{\infty} dv \,v^{2 \gamma-1}
K_0(v) \int_v^{\infty} dz \, q(z).  
\label{eq:integ_gamma}
\end{equation}

We first consider the parameter regime $\gamma < 1$. Apart from $\gamma=1$ the other ``easy" case is $\gamma=1/2$ where we find
\begin{equation}
P_{2,2}^{(1)}(\gamma=1/2)=\frac{1}{2 \pi}+\frac{5}{8} \approx 0.784155.
\end{equation}  
In fact, evaluation of such integrals with packages like Mathematica,
even in this case,  return unevaluated hypergeometric
functions. However, in this case, one can simplify the integrals
directly (hypergeometric functions encountered do not seem to have
known identities, at least to our knowledge). For instance, in the evaluation with the Bessel distribution part of the convolution, one needs
\begin{equation}
\frac{1}{\pi^2}\int_0^{\infty} \int_0^{\infty} K_0(v) K_0(v+z) \, dv\, dz = \dfrac{1}{2 \pi^2} \int_0^{\infty} \dfrac{u \, du}{\sinh u}= \dfrac{1}{8},
\end{equation}
which is obtained by using the integral representation of the Bessel function of zeroth order and simplifying. The integral with the first term of $q(z)$ is more easily evaluated to $1/(2 \pi)$. Thus the probability is $1/2+1/(2 \pi)+1/8$ as stated above. Numerical evaluation of these integrals is also tricky as $\gamma$ decreases toward zero. 

For $\gamma > 1$, the distribution $P_{2,2}^{(1)}(\gamma)=(1/2)+
A_1(\gamma)+A_2(\gamma)$ where 
\begin{equation}
A_1(\gamma)=\dfrac{\sqrt{\pi}}{2^{4\gamma} \Gamma(\gamma)^2} \sum_{k=0}^{2 \gamma-1}\frac{1}{2^k k!} \dfrac{\Gamma(k+2 \gamma)^2}{\Gamma(k+2 \gamma +\frac{1}{2})}, \;\; A_2(\gamma)=\dfrac{\Gamma(\gamma+\frac{1}{2})^2}{4 \sqrt{\pi} \Gamma(\gamma)^2}  \sum_{k=0}^{ \gamma-1}\frac{1}{k!} \dfrac{\Gamma(k+ \gamma)^2}{\Gamma(k+2 \gamma +\frac{1}{2})}.
\label{A1A2}
\end{equation}
The sums above maybe written in terms of hypergeometric functions, but they do not appear to be simple expressions. To state some other exact values for the probability of real eigenvalues:
\begin{equation}
P^{(1)}_{2,2}(2)=\dfrac{10259}{15015}, \,\, P^{(1)}_{2,2}(3)=\dfrac{640561}{969969}.
\end{equation}
The sum $A_1(\gamma)$ for large $\gamma$ can be approximated by
\be
A_1(\gamma) \approx \frac{\sqrt{\pi} \gamma^2}{2^{4 \gamma} (\gamma!)^2} ~\sum_{k=0}^{2 \gamma} \frac{1}{2^\gamma}~\frac{1}{(2 \gamma+k)^{3/2}}~ \frac{(2 \gamma+k)!}{k!},
\ee
where we have used that the ratio $x!/ (x+\frac{1}{2})! \approx x^{-1/2}$ for large $x$. Employing the Stirling's approximation $s ! \approx \sqrt{2 \pi s} (s/e)^s$ in the above equation, and replacing the sum by an integral, we obtain 
\be
A_1(\gamma) \approx \frac{1}{4 \sqrt{\pi}} \int_0^{2 \gamma} \frac{dk}{\sqrt{k}} ~\frac{2 \gamma}{2 \gamma+k} \exp \left[2 \gamma ~g \left(\frac{k}{2 \gamma} \right) \right],
\label{intapp}
\ee
where 
\be
g(x)=\ln(1+x)+x \ln(1+x^{-1})-(1+x) \ln 2. 
\ee
On evaluating the integral in Eq.~(\ref{intapp}) by saddle point
method, we obtain $A_1(\gamma)=(1/8) \textrm{erf}(\sqrt{\gamma/2})$
which, for large $\gamma$, gives $1/8$.   
An analysis similar to above also shows the leading order correction
in $A_1(\gamma)$ to be $(16 \sqrt{2 \pi \gamma})^{-1}$, and
$A_2(\gamma)$ to be exponentially small in $\gamma$.


\section{Distribution of the product for the probability distribution Eq.~(\ref{bddef})}
\label{appC}

For the distribution $p_{0,\nu}(x)$ defined by Eq.~(\ref{bddef}), the distribution of the product is given by
\bea
p_K(z_K) &=& \left(\frac{1+\nu}{2} \right)^K \int_{-1}^1 dx_1 ... \int_{-1}^1 dx_K ~|x_1|^\nu ...|x_K|^\nu  ~
\delta(z_K- \prod_{i=1}^K x_i) \\
&=& \frac{1}{2} ~(1+\nu)^K~ \int_0^1 dx_1... \int_0^1 dx_K ~x_1^\nu ... x_K^\nu ~\delta(|z_K|- \prod_{i=1}^K x_i),
\eea
where the last equation follows on noting that one half of the $2^K$ cases corresponding to the sign of the set $\{ x_i \}$  contribute equally to positive $z_K$. On carrying out the integral over $x_K$, we have
\bea
p_K(z_K) &=& \frac{1}{2} ~(\nu+1)^K ~|z_K|^\nu \int_{|z_K|}^1 \frac{dx_1}{x_1} 
\int_{\frac{|z_K|}{x_1 ... x_{K-3}}}^1 \frac{dx_2}{x_2} ... \int_\frac{|z_K|}{x_1 ... x_{K-2}}^1 \frac{dx_{K-1}}{x_{K-1}} \\
&=& \frac{1}{2} ~\frac{(\nu+1)^K}{(K-1)!} ~|z_K|^\nu
~\left[\ln \left(\frac{1}{|z_K|} \right) \right]^{K-1} ~,~ |z_K| < 1.
\eea
The above result can be obtained by either transforming the problem of
the distribution of product to that of sum by writing $\ln x_i=y_i$,
or directly using Mellin transforms \cite{Springer1966}.



\bibliographystyle{apsrev}
\bibliography{prod0315.bib}

\end{document}